\documentclass[12pt]{iopart}
\usepackage{graphicx}  
\usepackage{dcolumn}   
\usepackage{bm}        
\usepackage{amssymb}   

\usepackage{color}
\begin{document}


\title{Spontaneous oscillations from turnover of an elastic contractile material}
\author{Kai Dierkes$^1$, Angughali Sumi$^1$, J\'er\^ome Solon$^1$, Guillaume Salbreux$^2$}
\address{$^1$ Cell and Developmental biology programme, Center for Genomic Regulation (CRG), Barcelona 08003, Spain}
\address{$^2$ Max Planck Institute for the Physics of Complex Systems,  N\"othnitzerstr. 38, 01187 Dresden, Germany}
\date{\today}

\begin{abstract}
Single and collective cellular oscillations involving the actomyosin cytoskeleton have been observed in numerous biological systems. We show here that a generic model of a contractile material, which is turning over and contracts against an elastic element, exhibits spontaneous oscillations. Such a model can thus account for shape oscillations observed in amnioserosa cells during dorsal closure of the {\it Drosophila} embryo. We investigate the collective dynamics of an ensemble of such oscillators and show that the relative contribution of viscous and friction losses yield different regimes of collective oscillations. Taking into account the diffusion of contractile elements, our theoretical framework predicts the appearance of traveling waves which might account for the propagation of actomyosin contractile waves observed during morphogenesis.
\end{abstract}


In the cellular cytoskeleton, myosin molecular motors bind to networks of actin filaments and exert forces. Inhomogeneities in contractile stresses result in flows and deformations that can occur at the cellular or tissue level \cite{Salbreux:2012uq}. A striking example of such deformations are cell shape oscillations that have been shown to be involved in the morphogenesis of biological tissues during the development of living organisms \cite{Levayer:2012fk}. In many examples, these oscillations occur jointly with cycles of accumulation and depolymerization of actin and myosin in the cell cortex, a thin layer of actomyosin located near the surface of the cell and concentrated on the apical side of tissues \cite{Salbreux:2012uq}. A hallmark of the observed oscillatory behaviour is that the actomyosin density oscillates approximates in phase opposition with the cell shape, in cells \cite{Sedzinski:2011kx, Kapustina:2013kx} as well as in tissues \cite{Martin:2009fk, Solon:2009uq}. More specifically, higher concentrations reached when the cell surface area reaches a minimum \cite{Sedzinski:2011kx,blanchard2010cytoskeletal}.

A fundamental property of the cell actomyosin cortex is its ability to turn over and to constantly renew its constituents at a certain rate. A recent study \cite{Sedzinski:2011kx} highlighted that cortical turnover is essential to account for shape oscillations observed during cytokinesis. 

\begin{figure}[h]
\begin{center}
\includegraphics[width=14cm]{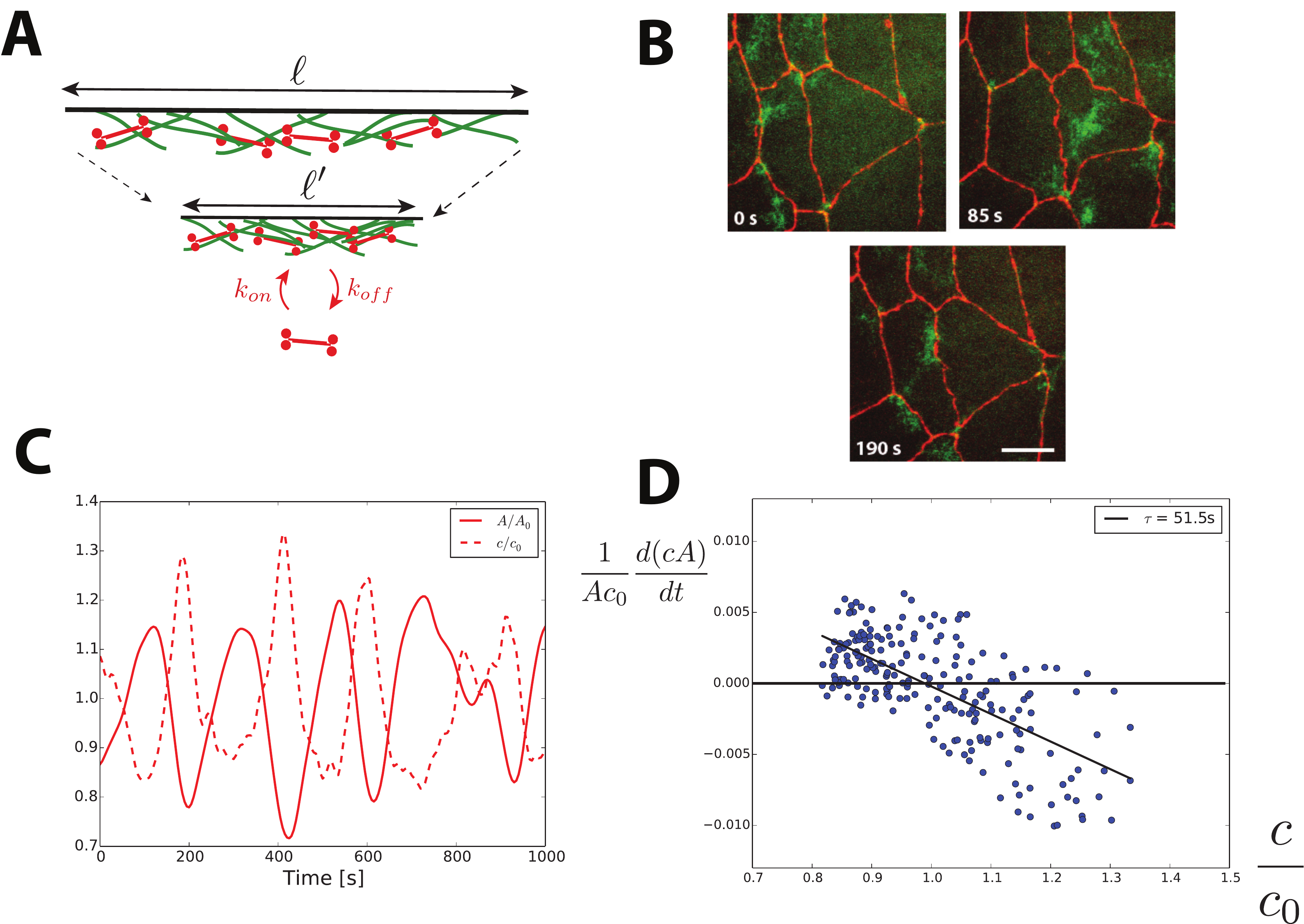}
\caption{\label{Figure1} A. Schematic of the apical cortex of amnioserosa cells. Myosin molecular motors accumulate when the network contracts, and exchange with a reservoir with rates $k_{on}$ and $k_{off}$. B. Snapshots of an oscillating cell during dorsal closure in the {\it Drosophila} embryo expressing Tomato E cadherin (red) and myosin, GFP-sqh (green). Scale bar, $10\mu$m.  C. Time series of the normalized area and average myosin concentration at the cell apical surface, for a representative cell in the amnioserosa.
D. Plot of the rate $\frac{1}{Ac_0}\frac{d(cA)}{dt}$ versus the normalized concentration change $\frac{c}{c_0}$ from the experimental measurement of amnioserosa cells. The points align on a line, in accordance with Eq. \ref{ChemicalEquation}. The slope of the line is related to the turnover timescale $\tau$. }
\end{center}
\end{figure}

We generalize here the model of the Ref. \cite{Sedzinski:2011kx} and  show that a minimal, generic description of actomyosin contractility predicts spontaneous oscillations and different patterns of collective oscillations observed in different biological tissues. The minimal ingredients to generate oscillations are (1) a contractile element whose constituents are turning over, (2) an elastic element and (3) a viscous damper, i.e. a  dissipative element. We think of the contractile element as representing a contractile network of concentration $c$ exchanging its material with a reservoir (Fig. \ref{Figure1}A), such that the concentration $c$ follows the following dynamical equation:
\begin{eqnarray}
\label{ChemicalEquation}
\frac{dc}{dt}&=&-\frac{1}{\tau}(c-c_0)-\frac{c}{l}\frac{dl}{dt}.
\end{eqnarray}
Here, the dynamics of the concentration $c$ arises from two effects: 1) binding and unbinding of contractile elements with rates $k_{\rm{on}}$ and $k_{\rm{off}} c$, such that $\frac{d c}{dt}=k_{\rm{on}}-k_{\rm{off}}c=\frac{1}{\tau}(c-c_0)$ with $\tau=\frac{1}{k_{\rm{off}}}$ and $c_0=\frac{k_{\rm{on}}}{k_{\rm{off}}}$, 2) in the absence of turnover, matter conservation imposes that the product $lc$ is constant, giving rise to the second term in eq. \ref{ChemicalEquation}.

To verify whether Eq. \ref{ChemicalEquation} can account for the dynamics of actomyosin cytoskeleton density in a tissue, we have measured the actomyosin intensity and cell surface area in cells of the amnioserosa during {\it Drosophila} dorsal closure (Fig. \ref{Figure1}B). Amnioserosa cells have been shown to exhibit oscillations of their apical surface with a period of $\sim 240s$ \cite{Solon:2009uq}. Measurements of the myosin intensity indicate that the myosin concentration oscillates approximately in phase opposition with the cell surface area (Fig. \ref{Figure1}C), with peaks in concentration slightly preceding minima in cell surface area. As predicted by Eq. 1, the rate $\frac{1}{A}\frac{d (cA)}{dt}$ is proportional to the myosin concentration $c$ during the oscillations of the apical surface area of the amnioserosa cell (see Fig. 1D). This result indicates that cortex turn-over and the changes in cell surface are sufficient to account for the dynamics of acto-myosin cortex concentration (Fig. 1D). The slope of the curve in Fig. 1D further allows to estimate the turnover timescale $\tau\simeq 50$s.

\begin{figure}[h]
\begin{center}
\includegraphics[width=14cm]{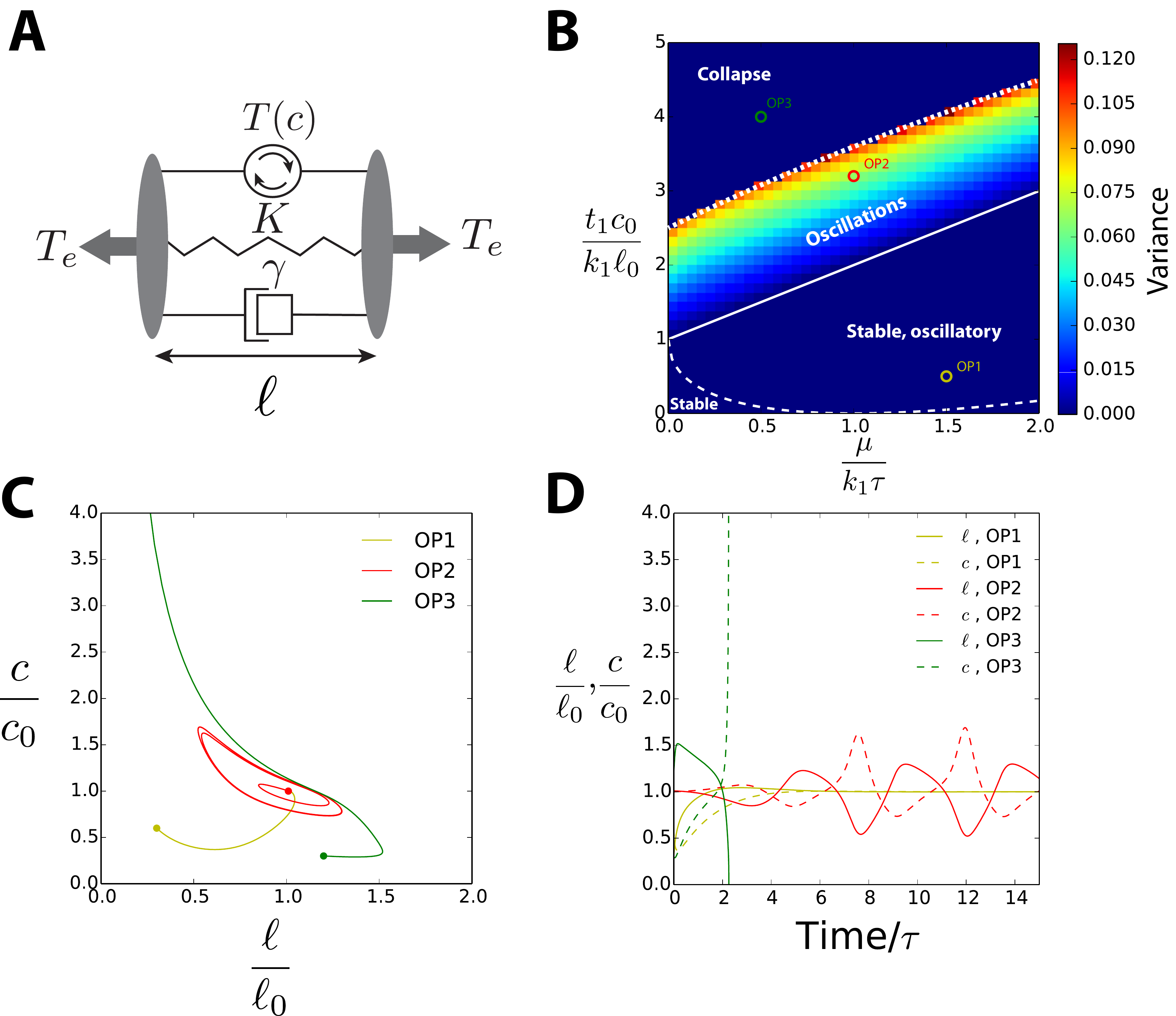}
\caption{\label{Figure2}A. Schematic of a minimal model of a mechanochemical oscillator. A spring and a dashpot are in parallel with a contractile material whose elements are turning over. The contractile material exerts a tension depending on the concentration $c$. B. Phase diagram for the behaviour of the unit shown in A, for $k_3/k_1=15$. The white line corresponds to a Hopf bifurcation to spontaneous oscillations. The dashed upper line is numerically evaluated and corresponds to a transition from a limit cycle to a collapse behaviour, where the length of the oscillators shrinks to $0$. The color code corresponds to the variance of the variance of the length $l(t)$ in the oscillatory steady-state. C, D Example trajectories of the unit in the three regions of the phase diagram, in a phase plot (C) and in a time series (D).  }
\end{center}
\end{figure}

Motivated by the observation that Eq. \ref{ChemicalEquation} is consistent with the dynamics in myosin concentration in the amnioserosa, we have investigated the properties of a simple model of cellular area deformation, assuming that the cellular contractile elements turn over according to Eq. \ref{ChemicalEquation}. We further assume that the mechanics of the actomyosin network can be described by a rheological element made of a viscous dashpot, an elastic spring and a contractile element (\ref{Figure2}A). The dynamical equation for the length of one contractile unit can then be written:
\begin{eqnarray}
\mu\frac{d l}{dt}&=&T_e-T(c) -K(l)\label{MechanicalEquation}
\end{eqnarray} 
The tension generated by the contractile unit is a function of the concentration of contractile generators $T(c)$, while $K(l)$ is an elastic restoring force which depends on the force-deformation relationship of the spring element. $T(c)$ and $K(l)$ are assumed to be monotonous functions of their respective arguments; $c$ and $l$. $\mu$ is a damping coefficient and $T_e$ an external tension opposing deformation of the unit (Fig. \ref{Figure2}A). 

At steady state the external tension balances the internal tension such that $T_e=T(c_0)+K(l_0)$. Expanding around the steady state, we write $T(c)=T(c_0)+t_1(c-c_0)+t_2 (c-c_0)^2+t_3 (c-c_0)^3$ and $K(l)=K(l_0)+k_1 (l-l_0)+k_2 (l-l_0)^2+k_3 (l-l_0)^3$. For the sake of simplicity, we assume that the response of the spring is symmetric to compression and extension, implying $k_2=0$. A linear stability analysis around the steady state shows that the system undergoes a Hopf bifurcation for 
\begin{equation}
\label{HopfBifurcation}
\frac{t_1 c_0}{ k_1 l_0}>1+\frac{\mu}{k_1\tau}.
\end{equation}
Matter conservation and the assumption that $T(c)$ increases monotonically with $c$ are responsible for the instability of the homogeneous state: contraction of the material leads to an increase of its density, leading to an even larger contractile force \cite{Salbreux:2007vn, Bois:2011ys, Hawkins:2011zr}. Turnover balances out this effect by restoring the reference concentration $c_0$. Above the bifurcation, a limit cycle appears and the systems undergoes spontaneous oscillations. The period of the oscillator at the bifurcation is $T=2\pi\sqrt{\tau \mu/k_1}$. The measured turnover time is $\tau\simeq50s$  (Fig. 1D). The timescale $\frac{\mu}{k_1}$ can be estimated from the relaxation of laser cuts in the apical actomyosin network of amnioserosa cells, and is of the order of  $90\pm 50s$ \cite{Solon:2009uq} . With these estimates, the period at threshold would be of the order of $T=425\pm150s$, i.e. comparable to the measured experimental period of cell oscillation $T=240s$. Moreover, the waveforms of simulated oscillations are very similar to experimental measurements (Fig. 1C), and the peak in concentration slightly precedes the minimum in surface area. Therefore, our very simple description can possibly account for cell oscillations {\it in vivo}.
Near the bifurcation line, the system can be brought to the normal form 
\begin{equation}
\frac{dz}{dt}=(r+i\omega)-(a+ib)|z^2|z
\end{equation} 
where $\omega=1/\sqrt{\tau \mu/k_1}$ is the pulsation at the transition, $r=\frac{1}{2}(\frac{t_1 c_0}{k_1 l_0}-1-\frac{\mu}{k_1 \tau})$ is a control parameter becoming positive above the transition, and $a$ and $b$ have lengthy expressions. 
The bifurcation type depends on the sign of $a$: assuming $k_2=t_3=0$ for simplicity, we find that for $3 \frac{k_3 l_0^2}{k_1} \left( \frac{ \mu}{k_1\tau}  + 1\right)>2(\frac{\mu}{k_1 \tau})^2+6 \frac{ \mu}{k_1 \tau} +4+t_2 \left(\frac{ \mu}{k_1 \tau}+7 \right)$, the transition is supercritical, and is subcritical otherwise. 

Further away from the bifurcation, we turn to numerical simulations to investigate the dynamics of the contractile unit.  In all numerical results shown in this manuscript, we have chosen to expand K to third order, with $k_2=0$ and $k_3$ large enough to ensure that the transition is supercritical and that the system reaches a limit cycle close to the bifurcation. We further choose $T(c)$ to be a linear function of the concentration $c$. The complete phase diagram is plotted in Fig. \ref{Figure2}B as a function of the reduced parameters $\frac{t_1}{k_1}$ and $\frac{\mu}{k_1 \tau}$. Further increasing the tension $t_1$ away from the Hopf bifurcation, a region appears where the system collapses to $l\rightarrow 0$ and $c\rightarrow\infty$ (see Fig. \ref{Figure2}C and D, green lines). The exact position of the transition separating stable oscillations from collapse depends on the nonlinearities in the functions $T(c)$ and $K(l)$ and  is numerically evaluated for $k_3/k_1=15$ in Fig. \ref{Figure2}B. The collapse occurs when the increase in tension as the contractile material accumulates overcomes the elastic resistance of the spring element. A sufficient condition for the system not to collapse is given by
$K(l=0)<T_e-T(c\rightarrow\infty)$: in that case $\frac{dl}{dt}(l\rightarrow 0)>0$ and the elastic resistance is large enough to counteract the tension exerted by the contractile material. This collapsing behavior could possibly be related to the delamination behavior of some cells whose apical area vanishes, as observed in the amnioserosa and other tissues \cite{Solon:2009uq, meghana2011integrin}.

\begin{figure}[h]
\begin{center}
\includegraphics[width=14cm]{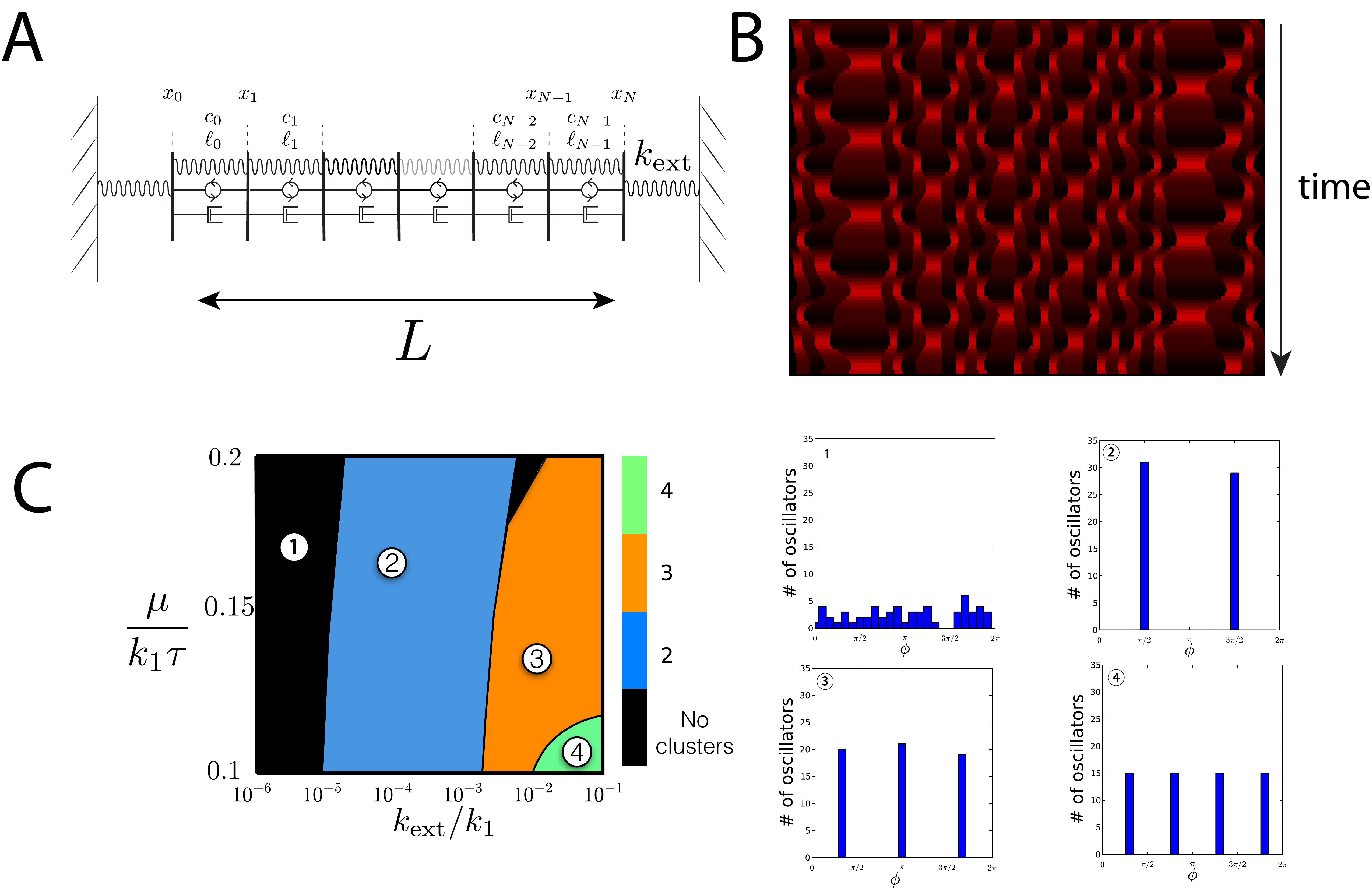}
\caption{\label{Figure3}A. Schematic of a chain of oscillators, connected to an external spring with resistance $k_{ext}$. B. Kymograph showing the time evolution of $N=40$ oscillators, with color coding for the concentration in each oscillator. The oscillators segregate in 3 synchronized clusters. C. Phase diagram of the collective oscillation modes for the chain shown in A. ($k_3/k_1=6$, $(t_1c_0)/(k_1l_0)=1.4$, $N=60$ oscillators). A transition appears from unsynchronized oscillations (black regions) to clustering of oscillators into 2 or more groups of synchronized oscillators. Lower panel, histograms of the phase distributions for parameters in different regions of the phase diagram.}
\end{center}
\end{figure}

\begin{figure}[h]
\begin{center}
\includegraphics[width=13cm]{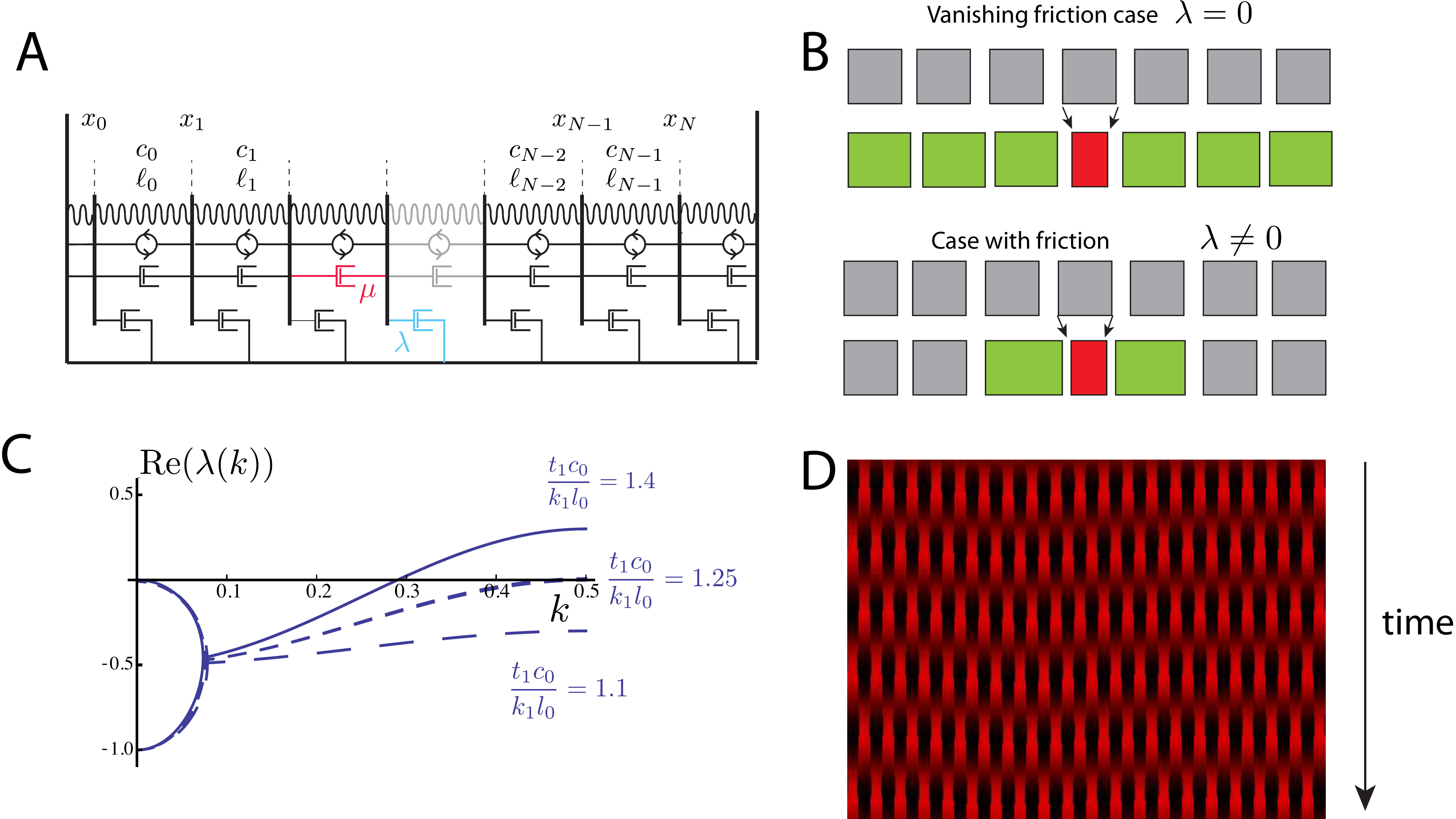}
\caption{\label{Figure4}A. Schematic of a periodic chain of oscillators, with linking points between the oscillators experiencing a friction relative to an external substrate. B. Schematic of the qualitative difference in mechanical coupling in the absence or presence of friction against an external substrate. In the absence of friction, contraction of one unit (red) drives expansion of all other units (green). In the presence of friction, a local contraction first initiates expansion of the nearest neighbors. C. Real parts of the eigenvalues of the linearized dynamics around the homogeneous state, as a function of the Fourier wave vector $k$, for $\frac{\tau k_1}{\mu}=1$ and $\frac{\mu}{\lambda}=0$. $k=\frac{1}{2}$ corresponds to a mode with nearest-neighbors oscillating in antiphase. D. Kymograph of a simulation of $N=40$ coupled oscillators with zero viscosity $\mu=0$, friction $\frac{\lambda}{k_1 \tau}=1.4$ and $\frac{t_1 c_0}{k_1 l_0}=1.4$. Color codes for the concentration $c$ in each oscillator.}
\end{center}
\end{figure}

In a tissue, oscillating cells are adjacent to each other and are mechanically coupled via adhesion proteins. We therefore now investigate the collective oscillations of a 1D periodic chain of $N$ of the mechanically coupled oscillating units (Fig. \ref{Figure3}A). We denote by $x_n$ the position of the boundary $n$ between two contractile units, and $l_n=x_{n+1}-x_n$ is the length of one unit. The boundaries between each unit could represent cell-cell interfaces in a tissue, or boundaries between independently contracting domains in the cytoskeleton. In addition to the viscous damper in each contractile unit, an additional friction force with friction coefficient $\alpha$ can oppose movement of the boundaries between the contractile units relative to a fixed external substrate (Fig. \ref{Figure4}A).

The mechanical equation for the position of one lattice point $x_n$ then reads
\begin{eqnarray}
\label{CoupledMechanicalEquation}
\lambda\frac{dx_n}{dt}&=&f_{n+1}-f_n\\
f_n&=&T_n-K(l_n)-\mu \frac{d l_n}{dt}
\end{eqnarray}
where $\lambda$ is a friction coefficient acting on the lattice points, and $f_n$ the total force exerted by the viscous, spring and contractile element in one unit. The dynamical equation for the concentration of one contractile unit is assumed to have the same form as Eq. \ref{ChemicalEquation}:
\begin{equation}
\label{CoupledChemicalEquation}
\frac{d c_n}{dt}=\frac{1}{\tau}(c_0-c_n)-\frac{c_n}{l_n}\frac{d c_n}{dt}
\end{equation}

   The uniform state $l_n=l_0$, $c_n=c_0$ is a steady state of the dynamical equations. We start by considering the case of vanishing external friction, $\lambda\rightarrow 0$. In that limit the total force in each mechanical unit is constant, $f_n=f$. The chain is assumed to be in contact with a spring of elasticity $k_{ext}$ at its boundaries, such that the total force is given by
\begin{equation}
f=k_{\rm{ext}}(L-N l_0)
\end{equation}
with $L$ the total length of the chain. The limit $k_{\rm{ext}}= 0$ corresponds to a free chain, while the limit $k_{ext}\rightarrow\infty$ correspond to a chain with a constrained total length $N l_0$.

 In the absence of friction relative to the substrate, the spatial distribution of the oscillators does not play a role, and the total force $f_n$ is conserved  along the chain. 
 In the limit of a free chain $k_{\rm{ext}}\rightarrow 0$, $f_n\rightarrow 0$, the oscillators are completely uncoupled and behave independently as the single unit oscillators considered in the first part of this manuscript. Taking non zero values of $k_{\rm{ext}}$ introduces a global coupling between the oscillators, as the contraction of one oscillator induce a reduction of the total length, and an increase in the total force along the chain which acts equally on every oscillator (Fig. \ref{Figure4}B). 
 To investigate the properties of the coupled oscillators, we perform a linear stability analysis of the regular lattice distribution by decomposing the motion of the lattice points in phonons, $\tilde{x}_k=\sum_n x_n e^{2\pi i n k}$. At linear order, the mode $k=0$ behaves effectively as a single unit with an effective elastic modulus $k_1+N k_{\rm{ext}}$, while all other modes $k\neq 0$ behave as a single unit with elastic modulus $k_1$. Therefore, as the normalized tension $t_1c_0/l_0 k_1$ is increased, the system undergoes a first Hopf bifurcation for all modes $k\neq0$, followed by a Hopf bifurcation for the mode $k=0$ describing the total length of the chain $L$.

Although all modes essentially uncouple at linear order, they can be coupled due to non-linearities. To investigate the resulting collective oscillations, we have numerically simulated a system of N oscillators (Fig. \ref{Figure3}B). To analyze the collective motion of the oscillators, we have numerically performed a phase reduction of the dynamical system, associating a phase $0<\phi<2\pi$ to points on the limit cycle.  We then analyze the distribution of phases of the oscillators (Fig. \ref{Figure3}C). We find that the chain of oscillators transitions from a uncoupled state with random phases for small external coupling $k_{\rm{ext}}$ to clustered states (Fig. \ref{Figure3}B). In a clustered state, the oscillators segregate into subpopulations which are fully synchronized, giving rise to several peaks in phase histograms (Fig. \ref{Figure3}B). The number of clusters appear to depend on the external spring modulus (Fig. \ref{Figure3}B), and we never find fully synchronized states.
Because the system is invariant by permutation of the oscillators, there is no spatial ordering of the clusters (Fig. \ref{Figure3}B). Besides, because of the nature of the mechanical coupling, contraction of one unit leads to expansion of other units; therefore, the coupling between units is repulsive. The frictionless chain is therefore in the class of globally coupled oscillators with a global repulsive coupling, which have been shown to elicit clustered states in a phase model of coupled oscillators, in accordance with our observations \cite{okuda1993variety}. 

We now consider the case of non-vanishing friction, $\lambda\neq 0$, where the oscillators are locally coupled. 
We consider here a periodic chain of constant length $L$. By performing a Fourier transform of eqs. \ref{CoupledMechanicalEquation}-\ref{CoupledChemicalEquation}, the linearized dynamics of phonons around the state $l_n=l_0$, $c_n=c_0$ is found to be 

\begin{eqnarray}
\frac{d}{dt}\left(\begin{array}{c}\frac{\tilde{x}_n}{l_0} \\\frac{\tilde{c}_n}{c_0}\end{array}\right)=M\left(\begin{array}{c}\frac{\tilde{x}_n}{l_0}\\ \frac{\tilde{c}_n}{c_0}\end{array}\right),\nonumber\\
M=\frac{k_1}{\lambda} \left(\begin{array}{cc}2 \frac{\cos(2\pi k)-1}{1- 2\frac{\mu}{\lambda}(\cos(2\pi k)-1)} &\frac{t_1 c_0}{k_1 l_0} \frac{e^{2\pi i k}-1}{1-2\frac{\mu}{\lambda}(\cos(2\pi k)-1)}\\-2\frac{(1-e^{-2\pi i k})\cos(2\pi k)-1}{1- 2\frac{\mu}{\lambda}(\cos(2\pi k)-1)} & -\frac{\lambda}{\tau k_1}+\frac{t_1 c_0}{k_1 l_0} \frac{2(1-\cos2\pi k)}{1-2\frac{\mu}{\lambda}(\cos(2\pi k)-1))}\end{array}\right).
\end{eqnarray}
 To study the stability of the homogeneous, non oscillating state, we have analysed the eigenvalues of the matrix $M$ as a function of the wave number $k$.  We find that the mode $k=\frac{1}{2}$ with nearest-neighbour antiphase oscillations is the first mode to be unstable as the tension $t_1$ is increased, for all parameters (Fig \ref{Figure4}B). The threshold tension for oscillation is given by
\begin{equation}
\frac{t_1 c_0}{k_1 l_0}>1+\frac{\mu}{\tau k_1} +\frac{\lambda}{4\tau k_1}
\end{equation}
which is larger than the single oscillation unit by a factor dependent on the friction $\lambda$. We indeed observe in numerical simulations that the units oscillate in phase opposition (Fig. \ref{Figure4}D). The major difference in oscillating patterns between the situation without friction and in the presence of friction can be understood as follows: in the absence of friction, the contraction of one unit results in a force acting equally on all other units, as force balance implies that the total force $f$ is conserved along the chain. With friction, contraction of one unit initially drives expansion of the nearest neighbors, and deformation is then propagated over time on units positioned at further distances (Fig. \ref{Figure4}B).

\begin{figure}[h]
\begin{center}
\includegraphics[width=16cm]{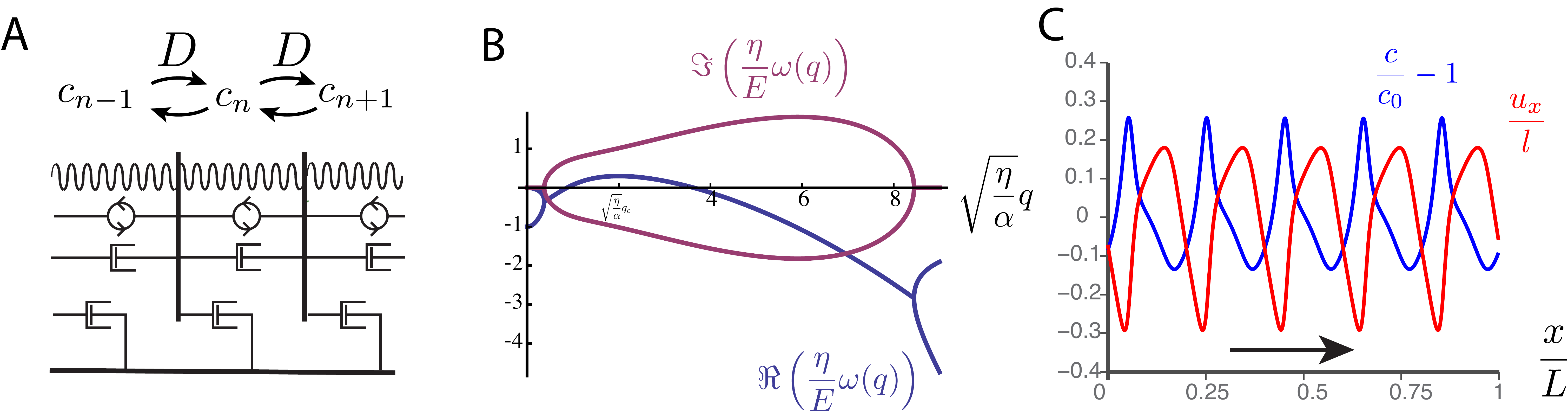}
\caption{\label{Figure5}A. Schematic for a chain of oscillators with diffusion of the concentration between different contractile units. B. Real and imaginary part of the eigenvalues of the linear stability matrix for the homogeneous state of Eqs. \ref{dtccontinuum}-\ref{forcebalancecontinuum} . For sufficiently large tension, a band of unstable modes appears around a critical wave vector $q_c$, with non zero imaginary value. Parameters are $\frac{D\alpha}{ E}=0.1$, $\frac{c_0 t'(c_0)}{E}=3.5$ and  $\frac{\eta}{\tau E}=1$. C. Snapshot of a result of a numerical simulation of Eqs. \ref{dtccontinuum}-\ref{forcebalancecontinuum} for $\frac{c_0 t'(c_0)}{E}=3.5$, $\frac{E_3}{E}=15$, $\frac{D\alpha}{ E}=0.1$, $\sqrt{\frac{\eta}{\alpha L^2}}=0.04$ with $L$ the system size, and $\frac{\eta}{\tau E}=1$. The simulation was performed with periodic boundary conditions. Traveling waves of concentration and deformation propagate in a direction depending on initial conditions. The black arrow indicates the direction of propagation.}
\end{center}
\end{figure}

We have considered so far oscillating units with independent concentrations $c$, which can represent connected cells separated by diffusion barriers. However, the actomyosin cortex of amnioserosa cells also exhibit a complex pulsing dynamics, with propagating actomyosin contractile waves appearing at the cell surface \cite{ma2009probing,blanchard2010cytoskeletal}. In the last part of this manuscript, we therefore turn to the case where the concentration in each oscillating unit diffuses across nearest-neighbor units. Such a description might be appropriate when cytoskeletal elements are free to diffuse, for instance to describe a continuous actin network (Fig. \ref{Figure5}A). The discrete model chain of oscillators can then be replaced by a continuous description in the limit $l_0\rightarrow 0, N\rightarrow\infty$:
\begin{eqnarray}
\label{ContinuumEquations}
\partial_t c&=&\frac{1}{\tau}(c_0-c)-\partial_x (v_x c)+D\partial_x^2 c\label{dtccontinuum}\\
f_x&=&E \partial_x u_{x}  +E_3 (\partial_x u_{x})^3+ \eta \partial_x v_{x}+t(c)\\
\partial_x f_x&=&\alpha v_x\label{forcebalancecontinuum}
\end{eqnarray}
where $E=k_1 l_0$, $E_3=k_3 l_0^3$ are two elastic moduli, $\eta=\mu l_0$ is the material viscosity and $\alpha=\lambda/l_0$ is a friction coefficient. $u_x(x)$ is a displacement field such that in the discrete limit $x_n= n l_0+u_x(n l_0)$ and $v_x=\frac{d u_x}{dt}$ is the associated velocity field. Eq. \ref{forcebalancecontinuum} corresponds to the force balance in one dimension for a system in contact with a substrate. Performing a spatial Fourier transform of eqs. \ref{dtccontinuum}-\ref{forcebalancecontinuum} allows to study the stability of the homogeneous state $c=c_0$, $u_x=0$. The homogeneous solution becomes unstable to an oscillatory periodic instability above the threshold tension
\begin{equation}
c_0 t'(c_0)>E+\frac{\eta}{\tau}+D\alpha+2\sqrt{\frac{D\alpha\eta}{\tau}}.
\end{equation} 
For non zero friction $\alpha\neq0$, the instability occurs at wavelength $\lambda=2\pi \left(\frac{D\tau \eta}{\alpha}\right)^{\frac{1}{4}}$ and with the pulsation $\omega=\sqrt{\frac{E}{\eta \tau}}$. At threshold, the system admits left or right traveling waves solutions with velocity $\left(\frac{DE^2}{\alpha\eta\tau}\right)^{\frac{1}{4}}$. Numerical simulations of Eqs \ref{dtccontinuum}-\ref{forcebalancecontinuum} with a spectral method indeed yield traveling wave solutions (Fig. \ref{Figure5}B). Traveling pulses of actomyosin have indeed been observed in the cortex of the amnioserosa \cite{ma2009probing,blanchard2010cytoskeletal} and might be explained by the traveling waves solution that we obtain here.

In summary, we have studied the individual and collective oscillations exhibited by a generic mechanochemical oscillator composed of a spring, a viscous dashpot and a contractile element whose units are turning over.
This minimal model recapitulates some of the essential features of cells oscillating {\it in vivo} and can explain the oscillations observed in amnioserosa cells during dorsal closure in the {\it Drosophila} embryo.  Mechanically coupling the oscillators introduces a repulsive coupling, as the contraction of one unit has to be balanced by expansion of the other units. We find that without friction against a substrate the oscillators are globally coupled, with a coupling strength that depends on the external resistance to deformation. Collective oscillation modes appear where oscillators are grouped in several fully synchronized subpopulations. In the other limit, i.e. for vanishing viscosity, friction against the substrate introduces nearest-neighbor couplings and favors anti phasic oscillations, ordering the oscillators in space. Therefore, analyzing the phase synchronization of oscillators in a tissue might yield informations of the relative values of tissue internal viscosity and friction relative to the substrate.
Including diffusion leads to the appearance of traveling waves in the system, which might correspond to waves observed in the apical actomyosin network of cells of the amnioserosa \cite{ma2009probing,blanchard2010cytoskeletal}.

In this manuscript, we chose a simple model for the evolution of the chemical concentration in the oscillators. One could imagine more complex dynamics for the concentrations of cytoskeletal elements in the cell, for instance coupled to the concentrations of proteins regulating the cytoskeleton. We have however made very basic assumptions on the mechanics of the oscillating units. We therefore think that our results on the collective oscillatory dynamics of an ensemble of mechanochemical oscillators will also apply to other examples of mechanically coupled oscillators, which may be common in other biological systems. The present work will therefore constitute a generic framework for further understanding of mechanochemical oscillations in biology.

\bibliographystyle{unsrt}
\bibliography{draft}

\end{document}